\documentclass[letterpaper,twocolumn,10pt]{article}
\usepackage{usenix-2020-09}

\usepackage{hyperref}

\usepackage{latexsym,amsmath,amsfonts,stmaryrd,mathtools}
\usepackage{amsthm}
\usepackage{algorithm, algorithmicx}
\usepackage[noend]{algpseudocode}
\usepackage{graphicx}
\usepackage{wrapfig}
\usepackage{listings, fancyvrb}
\usepackage{multirow}
\usepackage{comment}
\usepackage{xspace}
\usepackage{pifont}
\usepackage{parcolumns}
\usepackage{graphicx}
\usepackage{enumerate}
\usepackage{enumitem}
\usepackage{wrapfig}
\usepackage{tabularx}
\usepackage{titlecaps}
\usepackage{subcaption}
\usepackage{pgfplots}
\usepackage{booktabs}
\usepackage{upgreek}
\usepackage{enumitem}
\usepackage{soul}
\usepackage{mathtools}
\usepackage[scaled=0.9]{beramono}

\pgfplotsset{compat=newest}

\microtypecontext{spacing=nonfrench}

\makeatletter
\lst@Key{countblanklines}{true}[t]%
{\lstKV@SetIf{#1}\lst@ifcountblanklines}

\lst@AddToHook{OnEmptyLine}{%
	\lst@ifnumberblanklines\else%
	\lst@ifcountblanklines\else%
	\advance\c@lstnumber-\@ne\relax%
	\fi%
	\fi}
\makeatother

\usepackage{titlesec}
\usepackage{authblk}
\usepackage{svg}
\usepackage{colortbl}

\newcommand{\sysname}{DSig\xspace}

\newcommand{\us}{µs\xspace}
\newcommand{\wots}{W-OTS$^+$\xspace}
\newcommand{\haraka}{Haraka\xspace}
\newcommand{\blake}{BLAKE3\xspace}

\def\t{\textit}

\newenvironment{myitemize}{\begin{list}{\labelitemi}{
\setlength{\topsep}{0.5pt plus 0pt minus 0pt}
\setlength{\itemsep}{0pt plus 0pt minus 0pt}
\setlength{\parsep}{0pt plus 0pt minus 0pt}
\setlength{\parskip}{0pt plus 0pt minus 0pt}
}}{\end{list}}

\renewcommand{\paragraph}[1]{\smallskip\noindent\textbf{#1}}

\begin{document}

\title{\Large \bf \sysname: Breaking the Barrier of Signatures in Data Centers}

\author[1]{Marcos K. Aguilera}
\author[2]{Clément Burgelin}
\author[2]{Rachid Guerraoui}
\author[2]{Antoine Murat}
\author[3]{Athanasios Xygkis}
\author[4]{Igor Zablotchi}
\affil[1]{VMware Research Group}
\affil[2]{École Polytechnique Fédérale de Lausanne (EPFL)}
\affil[3]{Oracle Labs}
\affil[4]{Mysten Labs}

\renewcommand\Authfont{\rm}
\renewcommand\Affilfont{\itshape}
\setlength{\affilsep}{2em}

\maketitle

\begin{abstract}
Data centers increasingly host mutually distrustful
  users on shared infrastructure.
A powerful tool to safeguard such users are digital
  signatures.  
Digital signatures have revolutionized Internet-scale
  applications, but
current signatures are too slow for the growing genre of
  microsecond-scale systems
  in modern data centers.
We propose \sysname, 
the first digital signature system to achieve single-digit microsecond
  latency to sign, transmit, and verify signatures in
  data center systems.
\sysname is based on the observation 
that, in many data center applications, 
the signer of a message knows most of the time
who will verify its signature.
We introduce a new hybrid signature scheme that
  combines cheap
  single-use hash-based signatures verified in the foreground
  with traditional
  signatures pre-verified in the background.
Compared to prior state-of-the-art signatures,
  \sysname reduces signing time from 18.9 to 0.7\,\us and verification time
  from 35.6 to 5.1\,\us, while keeping signature transmission time below 2.5\,\us.
Moreover, \sysname achieves 2.5$\times$ higher signing throughput and 6.9$\times$ higher verification throughput than the state of the art.
We use \sysname to
  (a) bring auditability to 
two key-value stores (HERD and Redis) and a financial trading system (based on Liquibook)
for 86\% lower added latency than the state of the art, and
  (b) replace signatures in BFT broadcast and BFT replication, reducing their
  latency by 73\% and 69\%, respectively.
\end{abstract}

\section{Introduction}
\label{sec:intro}

\begin{figure}
    \smallskip
    \centering
    \includegraphics[width=\columnwidth]{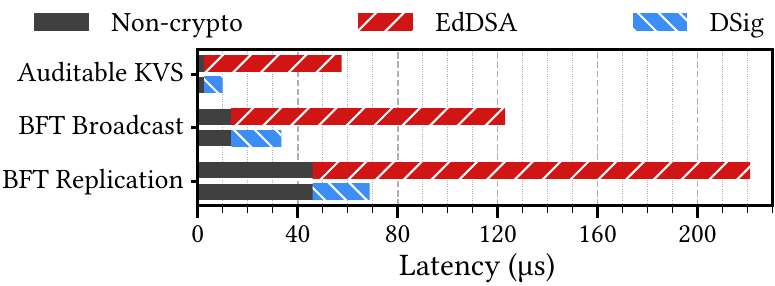}
    \caption{
    Median latency breakdown of an auditable key-value store (based on HERD~\cite{kalia2014using}, \S\ref{sec:apps}),
    a BFT broadcast primitive (CTB~\cite{ubft}, \S\ref{sec:apps}),
    and a BFT replication system (uBFT~\cite{ubft}, \S\ref{sec:apps})
    when processing small requests using either EdDSA~\cite{eddsa-rfc} (state of the art) or \sysname.
    \sysname reduces the cryptographic overhead by 86\%, 82\%, and 87\%, respectively, and reduces the overall latency by 83\%,  73\%, and 69\%, respectively.}
    \label{fig:intro-breakdown}
\end{figure}

Digital signatures
are used in many distributed protocols that
  have revolutionized the Internet
  through
  many
  use cases, such as
  enabling
  digital certificates~\cite{digitalcert},
  bootstrapping authentication protocols~\cite{tls, ssl3},
  securing and auditing transactions~\cite{
  nakamoto2008bitcoin, buterin2014ethereum},
  tolerating Byzantine failures~\cite{bft-smart, hyperledger-fabric, aguileraBGPXZ21},
  and
  verifying software authenticity~\cite{authenticode, sgx-explained}.
Signatures are irrefutable proofs that someone produced a
  message, and these
  proofs can be verified by third parties.
This property
  distinguishes signatures from
  message authentication codes (MACs)
  and authenticated symmetric
  encryption
  (e.g., SSL/TLS)~\cite{moderncryptobook}.
Today's signatures are however too expensive
for the growing genre of
  microsecond-scale systems in data
  centers.
Even the fastest signature scheme, EdDSA~\cite{eddsa-rfc,crypto-bench}, accounts for 79--95.6\%
  of the latency
  of applications such as auditable key-value stores,
  BFT broadcast, and BFT replication (Figure~\ref{fig:intro-breakdown}).

We propose \sysname,
the first digital signature
system to achieve single-digit microsecond performance for
  data centers.
A key insight underlying the design of \sysname is that,
  in many data center applications, signatures are issued and verified
  by parties known a priori in the common case,
  so signers and verifiers can
  exchange useful information beforehand and
  do part of
  the computation before knowing the messages
  to be signed,
  thereby reducing the latency of subsequent signature generation and verification.
We use this observation to introduce
a new hybrid online-offline
  signature system~\cite{even1990line}.
Hybrid schemes
  combine a traditional
  signature scheme that is slow but
  can sign many messages,
   with a hash-based signature scheme (HBSS) that is fast
   but
   can sign only one or a few messages.
The traditional signature is used to
  validate a batch of HBSS key pairs, each of which
  signs one or a few messages.
Hybrid signatures have been studied extensively in
  theory, but practical work has focused only on
 improving throughput
 for
  low-compute devices with limited bandwidth, on low-security signatures, or on tiny messages (\S\ref{sec:related}).

Using hybrid signatures to achieve low latency and high throughput in
  data centers poses a
  number of challenges.
First, the traditional part of hybrid signatures
  is compute-heavy and can impact latency.
Second, hybrid signatures involve frequent key pair generation, which can exhaust compute resources
  and impact throughput.
Third,
the actual performance of hybrid signatures
  deviates from their theoretical models, as
  real performance requires careful consideration of
  microarchitectural effects (e.g., caching, prefetching)
  rather than simply
  the amount of computation.
Fourth, to perform well, hybrid signatures need to be configured with an appropriate HBSS
  whose thousands of parameter combinations provide
  complex trade-offs between
  number of hash computations,
  signature size,
  and
  frequency of key pair generation;
  most parameter choices exceed
  our performance goals, the available network bandwidth,
  the computational resources, or all of the above.

We address
these challenges as follows.
First, we
  use hints about who will verify a message
  in the common case
  to pre-process the compute-heavy traditional signatures.
Second, we use traditional signatures to sign and verify batches of HBSS public keys, thus amortizing the cost of their authentication, while hiding the latency introduced by batching from the critical path.
Third, we study the
  real performance of
  HBSSs
  to determine the best schemes to use, and we discover
  non-intuitive cases where
  fewer hash computations
  actually harm performance.
Fourth, we identify two promising HBSSs to use in \sysname, \wots~\cite{wots-plus} and HORS~\cite{hors},
  and we explore their
  parameters in depth
  to understand
  how they affect latency, throughput, and resource usage;
  we give a recommended configuration of \sysname that strikes
  a good trade-off.

\begin{table}
\smallskip
\caption{
Comparison of EdDSA and \sysname in terms of
  latency to sign, transmit (tx) and verify;
  per-core sign and verify throughput;
  signature size; and background network traffic per signature with a single verifier.
 }
\label{table:config-overview}
\centering
\small
\setlength{\tabcolsep}{4pt}

\begin{tabular}{lccccccc}
\multirow{2}{*}{\textbf{}} & \multicolumn{3}{c}{\textbf{Latency (\us)}} & \multicolumn{2}{c}{\textbf{Tput (Kops)}} & \textbf{Sig size} & \textbf{Bg Net} \\
& Sign & Tx & Verify & Sign & Verify & \textbf{(B)} & \textbf{(B/Sig)} \\
\hline
EdDSA    & 18.9 & 1.1 & 35.6 &  53 &  28 &    64 &   0 \\
\sysname &  0.7 & 2.0 &  5.1 & 131 & 193 & 1,584 &  33 \\
\hline

\end{tabular}
\end{table}

We integrate \sysname with five
applications:
  two key-value stores (HERD~\cite{kalia2014using} and
      Redis~\cite{redis}),
  a financial trading system (based on
  Liquibook~\cite{liquibook}),
 a BFT broadcast primitive (CTB~\cite{ubft}),
  and a BFT replication system (uBFT~\cite{ubft}).
We use \sysname to
  provide
  auditability through a signed security log
in HERD, Redis, and Liquibook;
  and to replace the
   signatures
   used in
   CTB and uBFT
  to thwart
  Byzantine behavior.

We evaluate \sysname and its applications.
We find that \sysname
can sign, transmit, and verify a signature
in 7.7\,\us total,
  which is
  7.2$\times$ faster than EdDSA~\cite{eddsa-rfc}, the
  fastest traditional signature scheme~\cite{crypto-bench}
  (Table~\ref{table:config-overview}).
\sysname achieves 2.5${\times}$
  and 6.9${\times}$ higher throughput than EdDSA for
  signing and verifying.
\sysname benefits applications significantly.
In HERD, Redis, and Liquibook, \sysname brings
  auditability with an added latency
  of less than 8\,\us per operation,
  a reduction of 86\% in overhead compared to EdDSA.
In CTB, \sysname
   reduces the
   broadcast latency by 73\%,
   from 123\,\us  to 34\,\us.
In uBFT, \sysname
   reduces the
   replication latency by 69\%,
   from 221\,\us to 69\,\us.

The price for using \sysname
  is as follows.
First, to get the best performance,
 \sysname needs a priori knowledge of
  where and when signatures are verified
  (\sysname still works
  without such knowledge, but it is slower).
Second, \sysname requires
extra bandwidth and space to transmit and store its larger $\approx$1.5\,KiB signatures.
This
  is a small cost
  in data-center
  systems,
  which have
  low-latency high-bandwidth networks and
  abundant
  storage,
  but \sysname could be ill-suited for other
  settings, such as some wide-area systems.

In summary, our contributions are the following:
\begin{myitemize}
    \item We propose \sysname, a new digital signature system targeted at microsecond-scale
          applications in data centers.
          \sysname combines hash-based signatures,
          traditional signatures, and novel techniques to
          reduce latency in the critical path while achieving high throughput.
    \item We analyze and evaluate \sysname's large parameter space for low latency at high throughput, and identify a configuration that best fits most scenarios.
    \item We implement \sysname and integrate it into several applications:
          two key-value stores, a financial trading system, BFT broadcast, and a BFT replication system.
    \item We evaluate \sysname and its applications. \sysname significantly improves
          signature performance compared to
          EdDSA, the state of the art.
          These enhancements directly benefit the applications by
          providing better end-to-end latency and throughput, and by bringing auditability to the microsecond scale.
\end{myitemize}

\sysname is open source, available at
\url{https://github.com/LPD-EPFL/dsig}.

\section{Setting and Goals}
\label{sec:setting}
\label{sec:threat-model}

\paragraph{Setting.}
We target
   microsecond-scale applications~\cite{mukiller,putting-mu-back-in-muservices,shenango,ubft,aguilera2020microsecond, ukharon,
 zygos,arachne,ix,rpcvalet}
   with a few tens of servers within the same data center---a
   scale that addresses
  the computing needs of many enterprises.
These systems have
  a network with low latency (${\approx}1$\,\us)
  and high bandwidth (100s of Gbps or
  even Tbps~\cite{mellanox-brochure}).

\paragraph{Goals.}
Our goal is to achieve faster
  digital signatures
  to broaden
  their usability.
We do not seek general-purpose signatures
 for all settings
  (wide area networks, mobile networks, embedded
  systems), but rather seek schemes that provide
  the right trade-offs in
  modern data centers.
We seek signatures that provide the
  industry-standard level of security (128 bits).

Digital signatures are
  important because
  they are \emph{transferable}:
  if Alice signs a message to Bob, Bob can prove to
  Carol that Alice indeed signed it (\S\ref{sec:background}).
This property makes signatures more powerful than
  mechanisms such as
  SSL/TLS or MACs,
  which provide only symmetric authenticated
  channels between Alice and Bob~\cite{moderncryptobook}, which
  do not suffice for the applications
  we consider (\S\ref{sec:apps}).
Signatures help
  tackle distrustful parties
  in distributed protocols
  for a wide variety of use cases:
  securing transactions,
  enabling digital certificates,
  bootstrapping authentication of users and services~\cite{tls, ssl3},
  verifying software authenticity~\cite{authenticode, sgx-explained},
  providing integrity of
  audit logs~\cite{nakamoto2008bitcoin, buterin2014ethereum},
  tolerating Byzantine failures~\cite{bft-smart, hyperledger-fabric, aguileraBGPXZ21},
  etc.
Many of these use cases
  apply to microsecond-scale
  applications, as such systems
  increasingly bring together mutually distrustful users on shared infrastructure~\cite{synopsis-aws, illumina-aws}.
For example,
  microservices can benefit
  from Byzantine fault
  tolerance~\cite{ubft};
  signed transactions can provide
  auditability in high-frequency trading
  systems; and
  signed logs can provide
  a legal trail in high-stakes
  settings.

\paragraph{Threat model.}
Our threat model is standard for digital signatures~\cite{moderncryptobook}.
Malicious entities
can intercept, store, inject, delay, and alter messages.
We assume the security of standard cryptography
building blocks:
   traditional
   digital signature schemes
   (Ed25519~\cite{eddsa-rfc}),
   hash-based signature schemes
  (\wots~\cite{wots-plus} and HORS~\cite{hors}),
  and cryptographic hashes
  (SHA256~\cite{sha2}, \haraka (v2)~\cite{haraka}, and \blake~\cite{blake3}).
\section{Background} \label{sec:background}

\subsection{Digital Signature Schemes}

A digital signature scheme (DSS) has
  a
  key pair
  consisting of
   a public key $PK$ and a secret key $SK$.
A signer $s$ uses $SK$ to sign a message $m$, producing
  a signature $\sigma$ for $m$.
The signature $\sigma$ allows a party who knows $PK$ (and knows that $PK$ belongs to $s$) to verify that $m$ was signed by $s$.

DSSs provide
\textit{authenticity}, \textit{integrity},
\textit{public verifiability} and \textit{non-repudiation}
of messages~\cite{moderncryptobook}.
Authenticity means that a party with a message and its signature can verify the identity of the message's signer.
Integrity means that the party can verify that the message
  matches the message that was signed.
Public verifiability means that only $m$, $\sigma$, and $PK$ are needed to verify the authenticity and integrity of $m$.
Signatures are transferable: a party can use $\sigma$ and $m$ to convince anyone who
knows $PK$ that $m$ is authentic (and typically $PK$ is
published, so everyone knows $PK$).
This property differentiates digital signatures from other mechanisms,
such as message authentication codes (MACs), vectors of MACs,
authenticated channels (e.g., SSL/TLS),
and symmetric encryption (e.g., AES).
Non-repudiation means that $s$ cannot deny the signing of $m$ once its signature $\sigma$ is known.
Non-repudiation implies that signatures are non-forgeable:
  without knowing $SK$, it is computationally infeasible to produce a signature $\sigma$ which passes verification with $PK$.

\subsection{The Cryptographic Barrier}
After DSSs were proposed~\cite{diffie-hellman-public-crypto},
many schemes
followed:
RSA~\cite{rsa}, ECDSA~\cite{ecdsa}, EdDSA~\cite{eddsa-rfc}, etc.
These schemes rely on the hardness of certain problems
   (factoring, discrete logarithms) under sophisticated
   arithmetic (e.g., modular on elliptic curves).
They seek to provide strong security
  and small time to sign and verify.
For example, the state-of-the-art
128-bit-secure EdDSA
takes 19\,\us to sign and 36\,\us
to verify a small message on modern CPUs (Table~\ref{table:config-overview}).

State-of-the-art DSSs are
  too slow for modern data
  centers: even the fastest schemes
  are an order of magnitude
  slower than network latencies~\cite{crypto-bench}
  due to the use of sophisticated arithmetic,
  which consumes CPU and cannot be parallelized.
This slowness makes
 traditional DSSs prohibitive for distributed
  protocols, microservices, and applications
  that run at the microsecond scale,
  which need to check the signatures of messages
    before acting upon them.
For example, signature-based BFT protocols must check
  signatures before taking safety-critical steps such as computing a quorum intersection, voting, vouching for a message, deciding on a majority value, etc;
  similarly, auditable applications must check
  signatures before executing requests
  to provide auditability (\S\ref{sec:apps}).

Signing or verifying messages in batches can improve
the throughput of DSSs,
but batching further increases latency
and is thus ill-suited for latency-critical applications.

\subsection{Hash-Based and Hybrid Signatures}
\label{sec:hash-based-schemes}

\begin{figure}[b]
    \smallskip
    \centering
    \includegraphics[width=1.0\columnwidth]{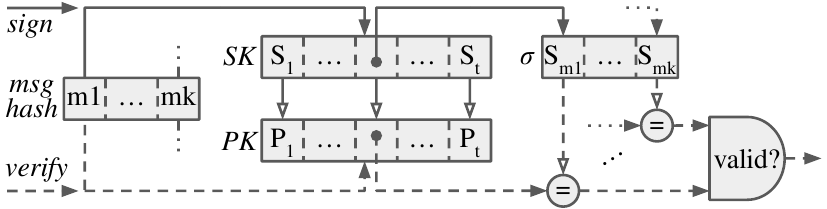}
    \caption{
    The HORS hash-based signature scheme.
                    Solid lines convey the path taken to sign a message,
                    while dashed lines convey
                    the path to verify a signature.
                    Hollow arrows indicate cryptographic hashes.}
    \label{fig:hb-schemes}
\end{figure}

Hash-based signature schemes (HBSSs)
were proposed
by Lamport~\cite{lamport1979constructing}.
They are DSSs that avoid advanced arithmetic
  by using only cryptographic hashes.
Hashes are advantageous because they can be computed quickly:
  modern algorithms (e.g., \haraka~\cite{haraka} and \blake~\cite{blake3}) can hash
  a small message in less than
  100\,ns on modern CPUs.
In some HBSSs (e.g.,
   HORS~\cite{hors}, \wots~\cite{wots-plus}), signature generation and verification execute at the
   microsecond scale,
   as they require few hash computations.

To explain HBSSs, we overview the HORS scheme
(Figure~\ref{fig:hb-schemes}),
which, whilst  simple, illustrates the key ideas of HBSSs.
The secret key $SK$ for signing is an array of $t$ random secrets
($t$ is a parameter), while
  the public key $PK$ for verifying is the concatenation of
  the hash of each secret in $SK$.
To sign a message, the signer
  hashes it with a salt
  into a string $m$, splits $m$ into $k$ substrings
  ($k$ is a parameter),
  treats each substring as an index into $SK$, and
  concatenates the indexed secrets to obtain
  the signature $\sigma$.
A verifier hashes the
  message with the salt
  and uses the substrings to index
  into $PK$.
Then, the verifier hashes each secret in $\sigma$ and checks that they match the indexed elements of $PK$.
This scheme is secure because it is hard to
  (1)~find messages that index the same secrets or
  (2)~reveal secrets without being the signer.
Other fast HBSSs, such as \wots~\cite{wots-plus}, are similar to HORS in that
  they sign by revealing a subset of
  the private key; as a result, they are limited to signing one or a few messages.

To overcome this limitation and sign an unlimited
  number of messages,
hybrid signature schemes~\cite{even1990line}
combine HBSSs with traditional schemes.
To sign a message $m$, a hybrid scheme concatenates an HBSS signature on $m$
  with the HBSS public key signed using a traditional signature.
To verify a signature, the scheme verifies the HBSS
  signature of $m$ and the traditional signature of the HBSS
  public key.

\subsection{Challenges}
Hybrid signatures were studied extensively in
  theory, but their application focused either on improving throughput
  in low-compute low-bandwidth devices,
  or on low-security signatures,
  or
  on tiny messages with only a few bits
(\S\ref{sec:related}).
To use them
  in a high-performance data center setting, we must
  tackle several challenges.

\paragraph{Efficient signature verification.}
To verify a hybrid signature,
  we must check both its HBSS signature and its traditional signature.
Traditional signatures, however, have high verification latency.
We need new mechanisms to avoid the traditional
  signature verification in the critical path.

\paragraph{Frequent key generation.}
Because an HBSS key pair can be used only once or a few times, hybrid schemes
  need to frequently generate and sign new HBSS key pairs.
This can become a bottleneck as it impairs signature throughput and, ultimately, its latency.
We need new techniques to improve throughput
  while minimizing latency on the critical path.

\paragraph{Practical performance.}
We evaluate the performance of hybrid schemes
  and find that it
  does not match their theoretical analysis.
The latter is based on simple metrics, namely the
  size of signatures and the number of hash calculations
  in the critical path.
However, due to microarchitectural effects
(e.g., CPU cache size, prefetching),
optimal configurations in theory
  perform suboptimally in practice, and
  optimizations that target solely the theoretical
  metrics sometimes do not work.

\paragraph{Large parameter space.}
Hybrid signature schemes have many configuration options:
  the choice of the traditional scheme, choice of the HBSS,
  and their respective parameters.
As a result, we are faced with thousands of options that
  provide different
trade-offs in network bandwidth, computational resources, throughput, and latency characteristics.

\section{Design of \sysname}
\label{sec:design}

We present the architecture of \sysname (short for ``Data center Signatures''), highlighting
its extended interface and computing planes (\S\ref{sec:overview}).
We then describe \sysname's hybrid signature scheme (\S\ref{sec:algorithm}),
its security (\S\ref{sec:security}),
and throughput optimizations (\S\ref{sec:eddsa-batching}).

\begin{figure}
    \smallskip
    \centering
    \includegraphics[width=1.0\columnwidth]{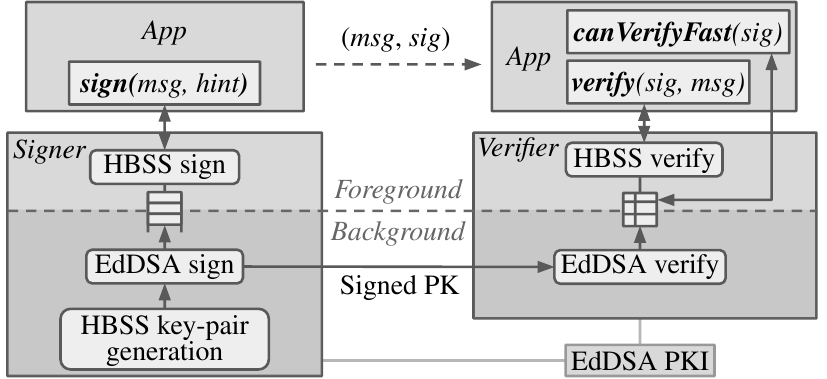}
    \caption{Architecture of \sysname.}
    \label{fig:arch}
\end{figure}

\subsection{Architecture}
\label{sec:overview}

Figure~\ref{fig:arch} depicts
the architecture
of \sysname.
Each process
has a public-private key pair of
  a traditional signature scheme, where
  the public key is made available to other parties
  via a public key infrastructure (PKI).
For the traditional signature scheme, we
  choose EdDSA~\cite{eddsa-rfc} because it is the fastest such
  scheme~\cite{crypto-bench}.
The PKI can be as simple as an administrator
  pre-installing the keys, or it can be
  a full-fledged system.

\sysname augments the interface of digital signatures (\t{sign} and \t{verify} functions)
  in two ways.
First, $\t{sign}$ takes a
  hint with the set of processes
  that
  will likely
  verify the signature.
The hint is optional: if omitted, it defaults to all known processes.
The hint does not restrict who can verify a signature---parties
  not indicated in the hint can still verify the signature, albeit at a lower
  performance.
Second, a new $\t{canVerifyFast}$ function
  returns whether verification of
  a given signature will be fast.
This function can mitigate denial-of-service attacks by letting
  applications prioritize the handling of fast signatures (\S\ref{sec:apps}).

Internally, \sysname has two planes:
\emph{foreground} and
\emph{background}.
The foreground plane provides the user library with a synchronous API
  to sign and verify messages,
  while the background plane asynchronously pre-generates and propagates
  new HBSS keys to be used by the foreground plane.
\sysname's general design can be used with a wide range of HBSSs
  (e.g., Lamport's~\cite{lamport1979constructing}, HORS~\cite{hors}, W-OTS~\cite{wots}, \wots~\cite{wots-plus}).
  We provide a specific recommendation based on an extensive performance study
  (\S\ref{sec:hbss-choice}).

\paragraph{Foreground plane.}
To sign a message, the signer uses a fresh
    HBSS key pair and returns to the application
    a \sysname signature, which includes
    the resulting HBSS signature
    and the
    corresponding HBSS public key signed with EdDSA.
Signing with EdDSA is slow, so the HBSS public key is
  pre-signed in the background plane.
On the other side, the verifier
    checks the authenticity of the message using the HBSS signature and the included HBSS public key.
The authenticity of the HBSS public key is checked by the background plane.

\paragraph{Background plane.}
The signer generates HBSS key pairs and EdDSA-signs them before forwarding them to the foreground plane.
It also sends the EdDSA-signed HBSS public keys to the background plane of the likely verifiers.
The latter verifies the authenticity of the HBSS public keys.

The background plane hides the latency of two
  slow steps:
(1) HBSS key pair generation, and
(2) EdDSA-signing and EdDSA-verifying the HBSS public keys.

Note that \sysname preserves the transferability of signatures irrespective of the
background plane.
Because \sysname hybrid signatures are self-standing
 (as they include the EdDSA-signed HBSS public key),
 the only requirement for signature verification is knowledge of the signer's EdDSA public key.
The background plane merely boosts performance
  when a hint is correct,
  by letting a verifier
  pre-check an HBSS public key
  before it receives a signature that includes it.

\subsection{Signing and Verifying in \sysname}\label{sec:algorithm}

\def\ContinueLineNumber{\lstset{firstnumber=last}}
\def\StartLineAt#1{\lstset{firstnumber=#1}}
\let\numberLineAt\StartLineAt

\begin{lstlisting}[float=t, caption={Signers' Pseudocode},label={alg:signer}]
@\textit{\# Signer's setup}@
verifier_groups = { ... } @\textit{\# Provided}@ @\label{alg:signer:verifier-groups}@
for group in verifier_groups:@\label{alg:signer:init-queues-begin}@
  signed_keypairs[group] = Queue()@\label{alg:signer:queue}\label{alg:signer:init-queues-end}@

@\textit{\# Signer's background plane}@
whenever @$\exists$@ group | signed_keypairs[group].size < @$S$@:@\label{alg:signer:while}@
  sk, pk = hbss.generate_keypair()@\label{alg:signer:gen_kp}@
  pk@$_\sigma$@ = {pk: pk, sig: eddsa.sign(pk)}@\label{alg:signer:sign_kp}@
  multicast <HBSS_PUBKEY, pk@$_\sigma$@> to group@\label{alg:signer:bcast_kp}@
  signed_keypairs[group].push((sk, pk@$_\sigma$@))@\label{alg:signer:push_kp}@
  
@\textit{\# Signer's foreground plane}@
def sign(msg, hint):
  group = @\textit{smallest group containing }@hint @\label{alg:signer:determine-group}@
  sk, pk@$_\sigma$@ = signed_keypairs[group].pop()@\label{alg:signer:extract_kp}@
  hbss_sig = hbss.sign(msg, sk)@\label{alg:signer:hbss_sign}@ 
  return <SIG, hbss_sig, pk@$_\sigma$@>@\label{alg:signer:return_sig}@
\end{lstlisting}

The logic of a \sysname signer is shown in
Algorithm~\ref{alg:signer}.
Each signer is configured with a list of \emph{verifier groups}---groups of processes that are likely to verify the same signatures on their critical path
  (line~\ref{alg:signer:verifier-groups}).
This list has a default group that contains all the processes in the system, but is otherwise application-dependent.
In the
  applications we examined (\S\ref{sec:apps}),
  the list was small and obvious
  (e.g., individual groups containing one process each).
Each verifier group is associated with a key-pair queue
(lines~\ref{alg:signer:init-queues-begin}--\ref{alg:signer:init-queues-end}).

In the background plane,
whenever a group's queue size is below a threshold
$S$ (line~\ref{alg:signer:while}),
the signer
generates a new HBSS key pair (line~\ref{alg:signer:gen_kp}),
and signs the public key using EdDSA (line~\ref{alg:signer:sign_kp}).
Empirically, we found that $S{=}512$ works well while consuming only 3\,MiB of memory per group.
Then, the signer multicasts
  the signed public key to the processes in the group (line~\ref{alg:signer:bcast_kp}).
The signer next appends the
private key with the EdDSA-signed public key
to the queue for consumption in the foreground plane (line~\ref{alg:signer:push_kp}).

To sign a message, the signer chooses the verifier group
  that matches the hint;
  if no group matches the hint,
  the signer picks the smallest group containing the hint (line~\ref{alg:signer:determine-group}).
Then, it gets a new HBSS key pair from this
  group's queue (line~\ref{alg:signer:extract_kp}).
Next,
the signer computes an HBSS signature of the message using the private key obtained from the queue
(line~\ref{alg:signer:hbss_sign}).
The \sysname signature comprises the HBSS signature of the message together with its EdDSA-signed HBSS public key (line~\ref{alg:signer:return_sig}).

\ContinueLineNumber

\begin{lstlisting}[float=t, caption={Verifiers' Pseudocode},label={alg:verifier}]
@\textit{\# Verifier's setup}@
verified_pks = Cache()

@\textit{\# Verifier's background plane}@
upon deliver <HBSS_PUBKEY, pk@$_\sigma$@> from process p:@\label{alg:verifier:upon_pk}@
  if eddsa.verify(pk@$_\sigma$@, eddsa_pk_of(p)):@\label{alg:verifier:verify_eddsa_bg}@
    verified_pks.add((pk@$_\sigma$@, p))@\label{alg:verifier:store_pk}@

@\textit{\# Verifier's foreground plane}@
def verify(msg, <SIG, hbss_sig, pk@$_\sigma$@>, p):@\label{alg:verifier:verify_sig}@
  if (pk@$_\sigma$@, p) not in verified_pks:@\label{alg:verifier:check_cache}@
    if not eddsa.verify(pk@$_\sigma$@, eddsa_pk_of(p)): @\textit{\# Slow}@@\label{alg:verifier:verify_eddsa_fg}@
      return false@\label{alg:verifier:verify_eddsa_return_false}@
  return hbss.verify(msg, hbss_sig, pk@$_\sigma$@.pk)@\label{alg:verifier:hbss_verify}@

def canVerifyFast(<SIG, _, pk@$_\sigma$@>, p): @\label{alg:verifier:can_verify_fast_def}@
  return (pk@$_\sigma$@, p) in verified_pks @\label{alg:verifier:can_verify_fast_body}@
\end{lstlisting}

The logic of a \sysname verifier is
shown in Algorithm~\ref{alg:verifier}.
In the background plane, the verifier receives
EdDSA-signed HBSS public keys (line~\ref{alg:verifier:upon_pk}), which
it verifies
(line~\ref{alg:verifier:verify_eddsa_bg}) and
stores
in a cache (line~\ref{alg:verifier:store_pk}).
In our applications, we found that having the cache store the latest $2{\times}S{=}1024$ HBSS public keys from each signer worked well while consuming only 100\,KiB of memory per signer.
In the foreground plane,
the verifier first consults its cache
  to see if it has a verified entry
  for the HBSS public key
  (line~\ref{alg:verifier:check_cache}).
If so,
the verifier proceeds
with checking the HBSS signature using the HBSS public key.
In this code path,
   the verifier checks signatures as fast as
   the underlying HBSS verify~(line~\ref{alg:verifier:hbss_verify}), which is fast.
Otherwise,
the verifier also checks
the EdDSA signature of the HBSS public key (line~\ref{alg:verifier:verify_eddsa_fg}),
so the verifier can operate even without the background plane.
The verifier's \textit{canVerifyFast}
function (\S\ref{sec:overview})
simply checks whether a signed HBSS public key has already been verified (lines \ref{alg:verifier:can_verify_fast_def}--\ref{alg:verifier:can_verify_fast_body}).

As with other signature schemes, \sysname
  can support key revocation through
  revocation lists that applications
  check prior to signing or verifying
  messages~\cite{moderncryptobook}.

\subsection{Security}\label{sec:security}

For preciseness of argument,
  we show the security of our recommended \sysname configuration, which uses \wots~\cite{wots-plus} as its underlying HBSS (\S\ref{sec:hbss-choice}).
Specifically, we show that this configuration of \sysname is
Existentially Unforgeable under Chosen-Message Attacks (EUF-CMA)~\cite{EUF-CMA} and that
it provides 128-bit security, which is safe by today's standards~\cite{too-much-crypto}.

\paragraph{EUF-CMA security.}
 We consider Chosen-Message Attacks (CMA)
 in which the attacker can query the target to sign arbitrary messages.
More precisely, we consider \textit{adaptive} CMA
 in which the attacker can query the target based on public key(s) and previously obtained signatures.
 We consider Existential Unforgeability (EUF),
 which means it should be computationally infeasible for an attacker to forge a signature on any message,
 except for messages that have already been signed by the target.

\paragraph{\sysname is as secure as its parts.}
To forge a signature in \sysname, an attacker must find a combination of message (not previously signed), \wots public key, EdDSA signature, and \wots signature that passes the \texttt{\small verify} function (Algorithm~\ref{alg:verifier}).
We assume that EdDSA provides EUF-CMA security, as
proved
by Brendel \textit{et al.}~\cite{eddsa-provable-sec}, and show how
\sysname's security reduces to the security of \wots:

\begin{enumerate}[leftmargin=15pt]
    \item \label{proof:parts:one} The verifier's background plane caches only correctly EdDSA-signed public keys
    (Alg.~\ref{alg:verifier} lines~\ref{alg:verifier:upon_pk}--\ref{alg:verifier:store_pk}).
    From the EUF-CMA security of EdDSA, and since
    a correct signer EdDSA-signs only its own public keys,
    (Alg.~\ref{alg:signer} lines~\ref{alg:signer:gen_kp}--\ref{alg:signer:sign_kp}),
    for any correct signer $s$, the verifier caches only the \wots public keys $s$ generates.

    \item If a public key is not cached,
    the verifier EdDSA-verifies it  on the critical path
    (Alg.~\ref{alg:verifier} lines~\ref{alg:verifier:check_cache}--\ref{alg:verifier:verify_eddsa_return_false}).
    As in (\ref{proof:parts:one}) above, for any correct signer $s$, this verification only succeeds for public keys $s$ generates.
    Thus, for any correct signer $s$, \texttt{\small verify} cannot return \textit{true} for public keys $s$ does not generate.

    \item Since, for any correct signer $s$, an attacker can only use a \wots public key generated by $s$,
    forging a \sysname signature reduces to forging a \wots signature.
\end{enumerate}

\paragraph{\wots with \haraka and \blake.}
H{\"{u}}lsing proved that \wots is EUF-CMA-secure when using a hash-function family that
is second-preimage resistant, undetectable, and one-way~\cite{wots-plus}.
Like SPHINCS+~\cite{sphincs-plus}, we pick the \haraka~\cite{haraka} hash-function family which satisfies those properties and relies on the battle-tested
AES round function.
Similarly to SPHINCS+, we reduce the signed messages to 128-bit digests by hashing them
salted with the \wots public key and a random nonce.
We do so using the well-established \blake~\cite{blake3} hashes.
Finally, we tune \wots's parameters to provide 128 bits of security when signing said 128-bit digests.
More precisely, we set the size of secrets and public key elements to 144 bits, which, together with a \wots depth of 4~(\S\ref{sec:hbss-choice}), provides a security level of 133.9 bits~\cite{wots-plus}.

\paragraph{\sysname's security level.}
Breaking \sysname can be reduced to breaking either EdDSA,
  \wots, \haraka, or \blake.
The EdDSA signature scheme Ed25519 provides 128-bit security
  under practical attacks~\cite{ed25519-web}, and our configuration of \wots provides 133-bit security. The security of both \haraka and BLAKE3 relies on well-studied components~\cite{too-much-crypto,haraka} and to date, no attack has compromised their security.

\subsection{Optimizing Throughput}
\label{sec:optimizations-bg-plane}

\sysname has a few throughput optimizations
  that do not significantly impact
  latency.

\paragraph{Speeding up key pair generation.}
Generating an HBSS key pair
  requires producing
  hundreds to thousands of
  secrets for the private key, and then hashing
  each secret for the public key.
To produce secrets quickly,
\sysname collects entropy
  from the hardware at startup to get a truly random 256-bit seed, which \sysname then salts with the key
  index and hashes using \blake
  to generate a digest with
  the size of the private key.
To produce the public key quickly, \sysname hashes
  the secrets using \haraka, which
  has a high-throughput implementation that optimizes
  instruction pipelining to compute multiple hashes efficiently.

\paragraph{Amortizing the cost of EdDSA signatures.}\label{sec:eddsa-batching}
EdDSA-signing every HBSS public key is slow and becomes a
  throughput bottleneck
   as each EdDSA sign-verify
  computation takes $\approx$55\,\us~\cite{crypto-bench}.
\sysname EdDSA-signs batches of HBSS public keys~\cite{rohatgi99}.
However, batching naively would increase the size of signatures,
since the entire batch of HBSS public keys should be included in every \sysname signature
to make their EdDSA signature self-standing.
\sysname addresses this issue by arranging the batch of HBSS public keys into a Merkle tree~\cite{merkle1978pkcrypto} and EdDSA-signing its root.
As a result, a \sysname signature is composed of an HBSS signature, an HBSS public key, a Merkle inclusion proof, and the EdDSA signature of the Merkle root.
The Merkle proof is a space-efficient way of proving that the included HBSS public key is part of the EdDSA-signed batch.
As Merkle proofs require collision-resistant hashes, we use (the efficient) \blake.
The space efficiency of Merkle proofs comes at the computational cost of generating and verifying them.
\sysname moves most of this cost to the background plane of both signers and verifiers, which precompute and cache the full Merkle tree associated with a batch.
Then, on the critical path, signing a key merely requires copying the subset of the tree that constitutes the Merkle proof,
  while verifying the Merkle proof consists of simple string comparisons.
The efficiency of this scheme depends on the batch size,
  which we determine in \S\ref{sec:eval-eddsa-batch}.

\paragraph{Speeding up bulk verification.}
Verifying many signatures with no assistance from the background plane (e.g., when checking an audit log) requires checking the same EdDSA signatures many times.
To speed up this process, EdDSA signatures verified in the foreground plane are cached.
A cache entry takes only ${\approx}33$ bytes,
  a tiny overhead, but saves $\approx$36\,\us
   of computation
  on our hardware~(Table~\ref{table:hwspecs}).

\paragraph{Reducing background bandwidth.}
Sending signed public keys both ahead of time
  and within signatures consumes
  significant networking bandwidth.
To nearly halve the bandwidth usage,
 \sysname batches, EdDSA-signs, and
  sends only
  \blake digests of the public keys in the
  background plane. This optimization requires computing the public key digest during signature verification, which adds only $\approx$1.3\,\us of latency.

\section{Choice of HBSS} \label{sec:hbss-choice}

\sysname can be used with many HBSSs, but its
  performance heavily depends on the actual
  HBSS used and how this HBSS is configured.
In this section, we explain which HBSS we choose for \sysname and how
  we configure it.

\subsection{HBSS Requirements} \label{sec:hbss-choice:req}

Our choice of HBSS is guided by the following requirements.

\paragraph{Security.}
To provide 128-bit security, \sysname needs an HBSS with
   the same or stronger security.

\paragraph{Low sign and verify latency.}
\sysname executes
HBSS \t{sign} and \t{verify} operations on the critical path.
These operations must have microsecond-scale latency.
This latency depends on the choice
  of the hash function, on the number of hashes they involve,
  and on microarchitectural effects.

\paragraph{Short signatures.}
At the microsecond scale, signatures cannot
  exceed a few KiB
  in length, as larger
  signatures incur significant transmission latency: we experimentally find that when
  sending small messages each extra KiB
  takes approximately an extra microsecond
  on a 100 Gbps network.
Furthermore, large signatures significantly increase
  the bandwidth consumption when applications send small messages.

\paragraph{Compressible public keys.}
Recall that \sysname signatures must include an HBSS public key in order to be self-standing (\S\ref{sec:algorithm}).
However, the combination of an HBSS signature and its public key can be in the KiB range, which is undesirable.
We thus seek HBSSs for which this combination can be compressed, leading to smaller \sysname signatures.

\paragraph{Lightweight key generation.}
HBSS key generation mainly involves hash computations, the number of which depends on the HBSS and can bottleneck \sysname's throughput. HBSS that use few hashes for key generation are thus desirable.

\subsection{Analysis} \label{sec:hbss-choice:analysis}

\begin{table}[!b]
\smallskip
\caption{Analytical comparison of a \sysname signature using either HORS or \wots as its HBSS for various configurations with EdDSA batches of 128 public keys.}
\label{table:hbss-metrics}
\centering
\small
\setlength{\tabcolsep}{4.5pt}
\begin{tabular}{c|cccc}

\multirow{2}{*}{\textbf{Conf}} &
\textbf{\# Critical} &
\textbf{Signature} &
\textbf{\# BG} &
\textbf{BG Traffic} \\

&
\textbf{Hashes} &
\textbf{Size (B)} &
\textbf{Hashes} &
\textbf{(B/Verifier)} \\
\hline

\multicolumn{5}{c}{\it Using HORS with factorized PKs} \\
\hline
k=8  & 8  & 8Mi  & 512Ki  & 33  \\
k=16 & 16 & 64Ki   & 4Ki    & 33  \\
k=32 & 32 & 8,552  & 512   & 33  \\
k=64 & 64 & 4,456  & 256   & 33  \\
\hline

\multicolumn{5}{c}{\it Using HORS with merklified PKs} \\
\hline
k=8  & 8  & 4,712  & 1Mi  & 8Mi \\
k=16 & 16 & 4,968  & 8Ki   & 64Ki  \\
k=32 & 32 & 5,480   & 1Ki  & 8Ki   \\
k=64 & 64 & 6,504   & 510   & 4Ki   \\
\hline

\multicolumn{5}{c}{\it Using \wots} \\
\hline
d=2  & $\approx$68  & 2,808 & 136  & 33  \\
d=4  & $\approx$102 & 1,584 & 204  & 33  \\
d=8  & $\approx$161 & 1,188  & 322  & 33  \\
d=16 & $\approx$263 & 990  & 525  & 33  \\
d=32 & $\approx$434 & 864  & 868  & 33  \\

\hline
\end{tabular}
\end{table}

HBSSs can be grouped in two classes: HBSSs with keys that can sign only one or a few messages~\cite{lamport1979constructing, hors, wots, wots-plus},
and HBSSs with keys that can sign
many messages~\cite{xmss, xmss-plus, sphincs, sphincs-plus, sphincs-plus-c, gravity-sphincs}.
Only the first class provides low latency
(the second focuses on quantum resistance).
Within that class, we focus on the fastest HBSSs: HORS~\cite{hors}
and \wots~\cite{wots-plus}.

\paragraph{HORS.}
Recall that a HORS signature reveals a subset of
  its private key secrets determined by the message being signed (\S\ref{sec:hash-based-schemes}).
Verifying a signature requires hashing the revealed secrets,
  checking that they match the public key,
  and checking that all the secrets mandated by the signed message
  were revealed.
HORS has two parameters: the number $k$ of secrets revealed in a signature
  and the number of times $r$ that a key pair can be used.
The size of HORS keys is proportional to $r$, so picking $r{\ge}2$ presents no benefits and we set $r{=}1$.
Smaller values of $k$, however, lead to fewer hash computations, and thus to
  lower latency in theory, but they require larger HORS public keys for the same security
  level.
Large HORS public keys require compression to fit our signature size budget.
We thus devise two compression techniques, described next.

\begin{figure}
    \centering
    \includegraphics[width=\columnwidth]{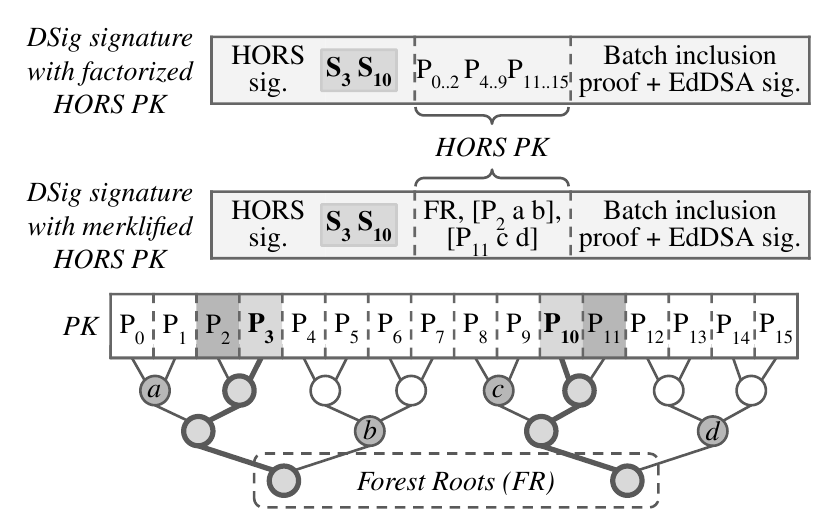}
    \caption{\sysname signatures when using HORS with either factorized or merklified public keys.
    For illustration purposes, we depict a toy configuration
    of HORS with 2-secret signatures and 16-element keys.}
    \label{fig:sig-layout-2}
\end{figure}

The first technique shortens
the embedded HORS
  public key by removing the elements that can be deduced from the HORS signature (top of Figure~\ref{fig:sig-layout-2}).
We analyze this approach
in the first part of Table~\ref{table:hbss-metrics}, which shows that configurations with few hashes ($k{<}32$) have large signatures (tens of KiB) that exceed our budget.

To use HORS signatures with small $k$ while staying within our signature size budget, we devise another compression technique
inspired by SPHINCS~\cite{sphincs}.
This technique is based on the observation that verifying a HORS signature merely requires checking that the few revealed secrets match the public key; knowledge of the full public key is unnecessary.
We enable such checks using Merkle inclusion proofs:
we arrange all public key elements in a Merkle forest, and EdDSA-sign its roots.
Such \sysname signatures replace their HORS public key with the forest roots and the inclusion proofs of the revealed secrets (bottom of Figure~\ref{fig:sig-layout-2}).
To avoid the overhead of checking Merkle proofs (i.e., computing around a hundred \blake hashes) on the critical path,
we use a latency-hiding technique similar to the one
  in \S\ref{sec:eddsa-batching}:
signers send their complete public keys
  ahead of time to verifiers (by disabling background bandwidth reduction (\S\ref{sec:optimizations-bg-plane}));
  verifiers
  precompute Merkle forests in their
  background plane, so Merkle proof verification on the critical path becomes mere string comparisons.
We analyze this approach in the second part of Table~\ref{table:hbss-metrics}, which shows that configurations with very few hashes ($k{\leq}16$) have tractable signature sizes, but come at the expense of significant background traffic and many background hashes.

\begin{figure}
    \centering
    \includegraphics[width=\columnwidth]{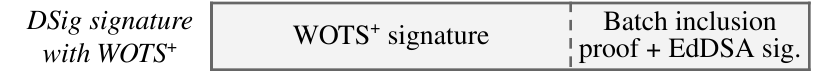}
    \caption{Layout of \sysname signatures when using \wots.}
    \label{fig:sig-layout}
\end{figure}

\paragraph{\wots.}
\wots differs from HORS in two main ways.
First, \wots secrets are hashed
   $d{-}1$ times to obtain the public key,
   where $d$ is a depth parameter.
Second, to sign, each secret is hashed a different number of times, as determined by
  the message to be signed, before being included in the signature.
We lower sign latency by caching these hashes upon computation
  of the public key so that signing reduces to string copying.
To verify a signature, we
  hash each element the required number of times to get to depth $d$, as determined
  by the signed message,
  and verify that the obtained results match the public key.
Note that \wots does not require embedding the public key in the \sysname signature (Figure~\ref{fig:sig-layout}).
A downside of \wots versus HORS is that \wots needs many more hashes
on the critical path.

We analyze \wots within \sysname in the last part of Table~\ref{table:hbss-metrics}, which
 shows that \wots configurations with small $d$ satisfy our requirements regarding signature size, background processing, and bandwidth consumption.
Moreover, although they all require more hashes than HORS on the critical path, their signatures are smaller than the smallest HORS ones.
Note that as $d$ gets bigger, the gain in signature size is outweighed by the drastic increase in hash computations.

\paragraph{Conclusion.} Our analysis points to three general configurations for further experimental evaluation: (1) HORS with factorized PK and $k$ close to 64, (2) HORS with merklified PK and $k$ close to 16, and (3) \wots with $d$ close to 4.

\subsection{Evaluation} \label{sec:hbss-choice:eval}
\begin{figure}
    \centering
    \includegraphics[width=\columnwidth]{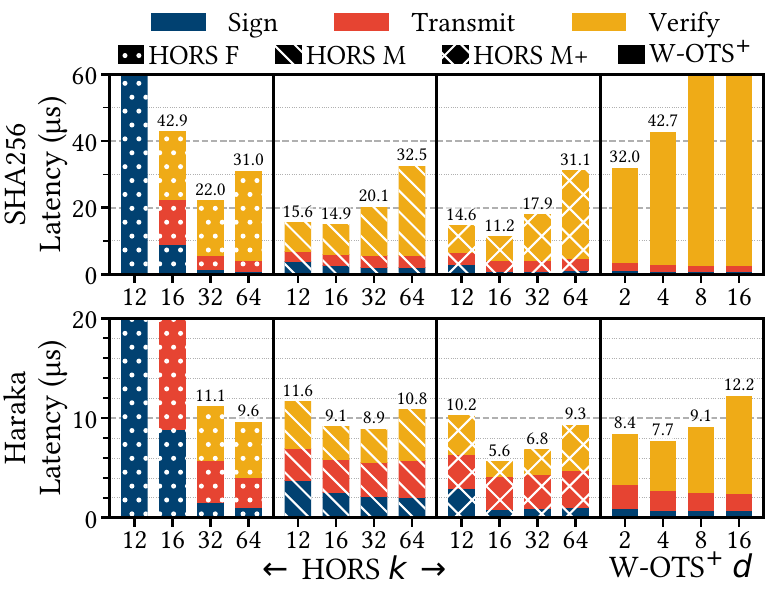}
    \caption{Sign-transmit-verify latency of \sysname for 8\,B messages using different
    hashing schemes (SHA256 (top) and \haraka (bottom)), and different
    HBSS configurations: HORS with factorized PK (HORS F), HORS with merklified PK (HORS M) and prefetched keys (HORS M+), and \wots.}
    \label{fig:hbss-vs}
\end{figure}

We measure the latency of signing an 8\,B message,
  transmitting it along with its \sysname signature, and verifying the signature for
  the sensible HBSS configurations presented in \S\ref{sec:hbss-choice:analysis}.
Our experimental setup is detailed in \S\ref{sec:evaluation}.
We consider three hash functions: SHA256~\cite{sha2} (the slowest),
  \blake~\cite{blake3}, and \haraka~\cite{haraka} (the fastest).
Figure~\ref{fig:hbss-vs} shows the results for SHA256 and \haraka (\blake stands in between).

When using \haraka (bottom of Figure~\ref{fig:hbss-vs}),
HORS with factorized public keys (denoted \emph{HORS F}) achieves its best end-to-end latency for $k{=}64$.
For $k{<}64$, its latency increases in spite of having fewer hashes on the critical path due to the transmission time of larger signatures.

Surprisingly, HORS with merklified public keys (\emph{HORS M}) signs
and verifies only marginally faster
despite its
far
lower number of hashes.
This disappointing performance is a microarchitectural effect of the size of the Merkle forests.
Indeed, for the assembly and verification of precomputed Merkle proofs to be fast, the relevant elements of their associated Merkle forest should be present in local cache (L1/L2).
However, CPU prefetching is inefficient due to the inherent randomness of Merkle proofs.

To demonstrate the benefit of having the Merkle proofs in the local cache,
we modify \sysname to prefetch public and private keys into the processor cache
right before signing and verifying.
The modified system (\emph{HORS M+}) achieves an end-to-end latency of as little as 5.6\,\us (signing in 0.9\,\us, verifying in 1.5\,\us and transmitting in 3.3\,\us) for $k{=}16$.
For smaller $k$, the keys do not entirely fit in the local cache, hence the performance degradation.

\wots achieves its best latency of 7.7\,\us for $d{=}4$ (signing in 0.7\,\us, verifying in 5.1\,\us and transmitting in 2.0\,\us) where it strikes a good balance between few hashes (low $d$) and short signatures (high $d$).
We omit prefetching \wots keys, as it has negligible latency benefit (not shown).

When using SHA256 (a slower hash function, top of Figure~\ref{fig:hbss-vs}), HORS with factorized public keys (HORS F) sees its verification time vastly increase for large $k$, while small $k$ still has long transmission time.
HORS with merklified public keys (HORS M) has better latency for smaller $k$,
    which differs from HORS M with \haraka.
Indeed, it is preferable to have fewer SHA256 computations (small $k$) than smaller Merkle forests (large $k$)
    since SHA256 is considerably more expensive than cache misses, which is not the case of  \haraka.
Finally, the large number of hashes (68 in expectation) of \wots makes it perform worse than the best configuration of each presented HORS variant.

\subsection{Recommended Configuration} \label{sec:hbss-choice:conclusion}
From our analytical (\S\ref{sec:hbss-choice:analysis}) and experimental (\S\ref{sec:hbss-choice:eval}) studies,
  we recommend using \wots with $d{=}4$ and \haraka.
This configuration offers single-digit sign-transmit-verify latency,
  tractable 1,584\,B signatures, and requires little
  background computation and networking.
Although HORS with merklified keys
  can achieve lower latency, it is too costly (in compute, bandwidth, and CPU cache pollution)
  and its superior performance requires modifying applications
  to prefetch keys into the
  processor cache, which can be impractical.
Note, however, that the choice of HBSS depends
  on the relative performance of hardware and software:
in the future, if cache misses become cheaper and
hashing becomes relatively more expensive,
  low-$k$ HORS configurations could be appealing.

\section{Applications} \label{sec:apps}

We apply \sysname to key-value stores, a financial
trading system, BFT broadcast, and BFT replication.

\paragraph{Key-value stores (HERD~\cite{kalia2014using} and Redis~\cite{redis}).}
State-of-the-art key-value stores provide microsecond-level performance and form
the backbone of many data center applications, microservices, and cloud services.
Key-value stores are used to keep
  security-sensitive information such as
  service configuration, session management data, cached queries,
  access control data, chat sessions, etc.
Yet, most key-value stores lack {\em auditability}---the ability for a third-party to check a log of all
operations (reads and writes) executed on the key-value store.
More precisely, in an auditable key-value store, the
  server keeps a log of executed operation such that, for
  any operation $\textit{op}$ in the log, the server can prove
  to a third party
  that $\textit{op}$'s client requested its execution.
For example, the third party may be a forensics specialist
  or a prosecutor, who wants proof that a client requested
  access or modification to some data.
The threat model is that clients
  may wish to bypass the audit (i.e., execute an operation
  undetected), while the server is honest.
The proofs provided by the server are operations signed
  by clients and the key-value store must ensure that (a) if an operation signed
by client $C$ is in the audit log, then it was
executed by the key-value store for client $C$, and
(b) if an operation is executed by the key-value store
for client $C$, then it appears in the audit log as
an operation signed by $C$.

To provide auditability, a key-value store
  requires all requests to be signed by clients
  and
  logged by the server.
The server must check the client signature before
  executing a request (otherwise a client could
  send a request with a bogus signature, which the
  server would execute without later being able to
  prove it),
  so traditional digital signatures
  significantly increase the latency of microsecond-scale key-value stores.

We use \sysname to add auditability to
  two key-value stores:
  HERD and Redis.
HERD is highly optimized for
  the RDMA networks present in data centers, while
  Redis is popular among web application developers.
HERD provides simple \texttt{\small GET} and \texttt{\small PUT} operations on key-value pairs,
  while Redis also provides higher-level operations on common data structures, such
  as lists, maps, sets, etc.
We modify both systems so that clients use \sysname to sign
all operations,
  and servers log and verify the signed operations before executing them.
This logging requires 1.5\,KiB of storage per operation due to \sysname's signatures.
Key-value stores have predictable signing and verifying processes:
clients simply set their signatures hints to the server process.
Logs can be persisted at the microsecond scale using persistent memory.
This is not currently implemented due to
  lack of
  hardware, but data from Yang
  \textit{et al.}~\cite{optane-benchmark} indicate that persistence would take less than 4\,\us, and
  this latency can be masked by storing in parallel with signature verification.
Vanilla
HERD takes $\approx$2.5\,\us
  to \texttt{\small GET} or \texttt{\small PUT} a key-value pair,
  while vanilla Redis takes $\approx$12\,\us.

\paragraph{Financial trading system (Liquibook~\cite{liquibook}).}
Liquibook provides an order-matching engine for
  financial trading---it
matches buy and sell limit orders from clients.
We consider a trading system built using Liquibook and
  RDMA for fast client-server communication.
We use \sysname to enhance the system and provide
  auditability, as we did for key-value stores.
Signature hints are identical to key-value stores.
Without auditability, the trading system takes $\approx$3.6\,\us
  to process orders,
  of which $\approx$2\,\us are spent on communication.

\paragraph{BFT broadcast (CTB~\cite{ubft}).}
Byzantine Fault Tolerance (BFT) is becoming more relevant in
  data centers, due to the need
  to tolerate failures beyond simple crashes~\cite{flash-memory-failures, dc-net-reliability, cores-that-dont-count, fail-slow-at-scale, gray-failure}.
Consistent broadcast is a core BFT primitive that prevents
  equivocation~\cite{rachid-book} and has many uses, as in
  money transfer~\cite{astro,The-Consensus-Number-of-a-Cryptocurrency}, consensus~\cite{fast-asynchronous-consensus, on-the-limited-power, revisiting-the-power}, multi-party computation~\cite{Asynchronous-MPC-non-equivocation} and decentralized learning~\cite{learning-in-the-jungle, robust-collaborative-learning} protocols.
We consider uBFT's state-of-the-art
implementation of Consistent Tail Broadcast (CTB) and replace its signatures with \sysname's to improve performance.
Signing hints are simple, as each signature is verified by all
processes running the protocol.

\paragraph{BFT replication (uBFT~\cite{ubft}).}
State machine replication (SMR) is the standard approach
for fault-tolerance~\cite{mencius, aguilera2020microsecond, bft-smart}.
We consider uBFT,
  a microsecond-scale BFT SMR system for data centers.
BFT SMR protocols, including uBFT, often employ signed messages
  to guard against Byzantine replicas.
uBFT recognizes the high cost of digital signatures and follows a fast/slow path approach. The fast path avoids signatures and has a latency of 5\,\us.
The slow path uses signatures, with a latency of ${\approx}$220\,\us.
The slow path is triggered even without Byzantine behavior (e.g., due to process slowness),
leading to latency fluctuations between its two modes of operation.
We use \sysname to replace the digital signatures in uBFT and improve its performance.
Signing hints are simple, as each signature is verified by all processes
  running the protocol.
Moreover, we use \sysname's DoS-mitigation mechanism (\S\ref{sec:design}) to prevent a malicious attack
from triggering superfluous EdDSA verifications.
More precisely, because uBFT is a quorum system, it can make progress with
  $n{-}f$ responses ($n$ is the number of replicas, $f$ is the maximum number that can be Byzantine).
We make a small modification to uBFT to use the \t{canVerifyFast} function
  to prioritize handling of messages that do not incur the EdDSA signature check.
  As a process gets at least $n{-}f$ messages from non-Byzantine processes, it ignores the slow-to-check messages from Byzantine players.

\section{Implementation}

Our implementation of \sysname has
 3,019 lines of C++17 (CLOC~\cite{cloc}).
We use our own implementation of HORS~\cite{hors} and \wots~\cite{wots-plus},
  the official implementations of \blake~\cite{blake3} and \haraka~\cite{haraka},
  and Dalek's implementation of EdDSA (Ed25519~\cite{dalek}).
\blake and Dalek's EdDSA use AVX2 for high performance.
We use uBFT's framework~\cite{ubft}, which provides fast
  point-to-point
  communication.

\begin{table}
    \smallskip
    \setlength{\tabcolsep}{2pt}
    \caption{Configuration details of machines.}
	\label{table:hwspecs}
\small
 \centering
	\begin{tabular}{cl}
	\toprule
\textbf{CPU}		& 2$\times$ 8c/16t Intel Xeon Gold 6244 @ 3.60GHz \\
\textbf{NIC/Switch}		& Mlnx CX-6 MT28908 / MSB7700 EDR 100 Gbps \\
\textbf{Software}  & Linux 5.4.0-167-generic /  Mlnx OFED 5.3-1.0.0.1 \\
	 \bottomrule
	\end{tabular}
\end{table}

\section{Evaluation} \label{sec:evaluation}

We evaluate the performance of \sysname
and verify its suitability as a microsecond-scale signature system.
We aim to answer the following:

\begin{myitemize}
    \item How do microsecond-scale applications that use signatures benefit from \sysname's low latency  (\S\ref{sec:eval-e2e-lat})?
    \item How does \sysname's signing and verification latency compare to traditional signatures (\S\ref{sec:evallat}, \S\ref{sec:msgsize})?
    \item What is the throughput of \sysname (\S\ref{sec:eval-tput})?
    \item How do \sysname's higher bandwidth requirements impact its scalability (\S\ref{sec:eval-scalability}, \S\ref{sec:eval-end-tput})?
    \item How do we set \sysname's EdDSA batch size
    (\S\ref{sec:eval-eddsa-batch})?
\end{myitemize}

\paragraph{Testbed.}
Our testbed is a cluster with 4 servers configured as shown in
  Table~\ref{table:hwspecs}.
The dual-socket machines have an RDMA NIC attached to the first socket.
Our experiments execute on cores of the first socket using local NUMA memory.
We accurately measure time using the TSC~\cite{tsc} via
\texttt{\small clock\_gettime} with the \texttt{\small CLOCK\_MONOTONIC} parameter.

\paragraph{\sysname configuration.}
We configure \sysname per \S\ref{sec:hbss-choice}: in all experiments, we use
  \wots with a depth $d{=}4$ as its HBSS.
We dedicate a single core to \sysname's background plane, which provides a high-enough throughput for our applications (\S\ref{sec:eval-tput}).
  Unless specified otherwise, we use an EdDSA batch size of 128 (\S\ref{sec:eval-eddsa-batch})
  and provide correct verifier hints when signing.

\paragraph{Baselines.}
We compare \sysname against two well-known signature libraries:
  Sodium~\cite{sodium} (written in C) and Dalek~\cite{dalek} (written in Rust).
Both
implement the EdDSA
signature scheme Ed25519---the fastest traditional scheme to date~\cite{crypto-bench}.

\subsection{Application Latency}\label{sec:eval-e2e-lat}

\begin{figure}
    \smallskip
    \centering
    \includegraphics[width=\columnwidth]{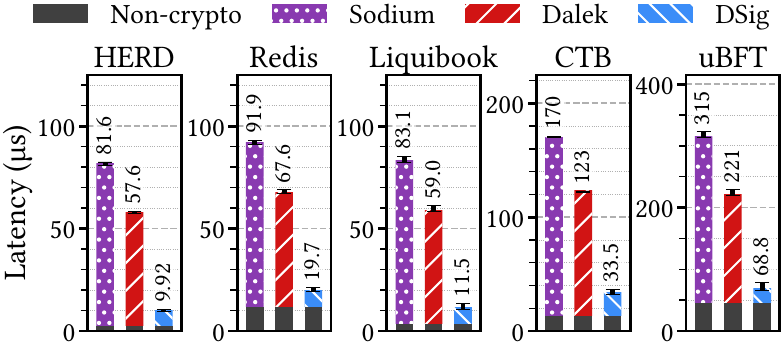}
    \caption{End-to-end latency of different applications using Sodium, Dalek or \sysname for signatures.
    Printed values show the median latency; whiskers show the 10th and 90th \%-iles.
}
    \label{fig:app-latency}
\end{figure}

We configure
applications with different signature schemes (Sodium, Dalek, \sysname)
  and measure their
  latency.
For the key-value stores, we use 16\,B keys and 32\,B values,
  20\% of \texttt{\small PUT} requests and 80\% of \texttt{\small GET}s, where 90\%  of \texttt{\small GET}s return a value.
For Liquibook, 50\% of the requests are \texttt{\small SELL}s and 50\% are \texttt{\small BUY}s.
For CTB, we broadcast 8\,B messages.
Finally, for uBFT, we consider SMR operations of 8\,B.
We issue $10,000$ requests one at a time to each application,
measure the
end-to-end latency,
and report the 10th-, 50th-, and
  90th-percentiles.

Figure~\ref{fig:app-latency} shows the results.
For the three applications on the left, \sysname provides auditability
for a small cost: an increase of less than 7.9\,\us in end-to-end latency.
Sodium and Dalek add $\approx$79\,\us and $\approx$55\,\us, respectively,
  which is 10$\times$ and 7.0$\times$ \sysname's overhead.
In CTB, replacing Sodium (resp. Dalek) with \sysname
reduces the median cryptographic overhead by
87\% (resp. 82\%),
and reduces the
median end-to-end latency by
80\% (resp. 73\%).
In uBFT, \sysname
reduces the median cryptographic overhead by
91\% (resp. 87\%),
and reduces the
median end-to-end latency by
78\% (resp. 69\%)
compared to Sodium (resp. Dalek).
\sysname provides similar latency gains at the 90th percentile.
In summary, across the tested applications, \sysname significantly reduces cryptographic overheads and improves latency over the state of the art.

\subsection{Latency of \sysname}
\label{sec:evallat}

We study the latency to sign a message, transmit a signature, and verify a signature
  using \sysname.
We also consider the latency of incorrectly hinted
  \sysname signatures for which
  EdDSA signatures are verified on the critical path.
  This represents the worst-case scenario for \sysname.
In each experiment, a process signs an 8\,B message and
  transmits the signed message
  to a second process, which verifies the signature.
We run each experiment 10,000 times for
  each signature scheme.
The signature transmission latency is the incremental cost of adding the signature to a message, computed as the difference between transmitting a message with and without a signature.
We estimate message transmission time as half of the round-trip
  time for ping-ponging the message.

\begin{figure}
    \smallskip
    \centering
    \includegraphics[width=\columnwidth]{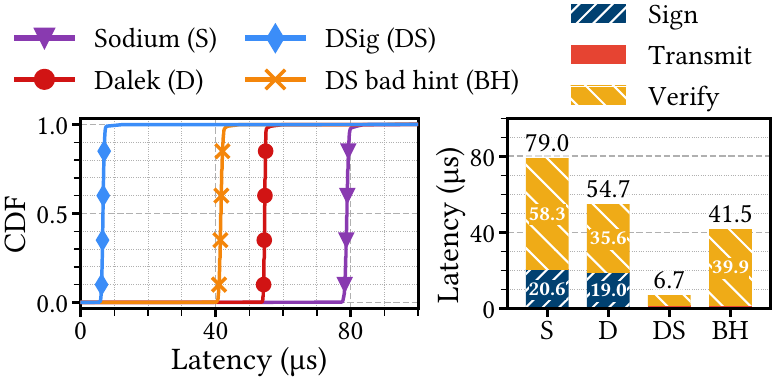}
    \caption{(Left) CDF of latency when signing, transmitting, and verifying the signature of 8\,B
    messages using Sodium, Dalek and \sysname with/without correct hints.
    (Right) Median latency breakdown. Transmission overhead is barely visible.}
    \label{fig:pony-vs-dalek-sodium}
\end{figure}

Figure~\ref{fig:pony-vs-dalek-sodium} shows the results.
All signature schemes have stable latency up to the 99.9th percentile.
Sodium and Dalek have similar signing median latency of 20.6\,\us and 18.9\,\us, respectively.
While Sodium verifies in 58.3\,\us, Dalek verifies in only 35.6\,\us (39\% faster)
thanks to the use of AVX2 instructions.
The (incremental) network latency is less than 100\,ns for both since their signatures are merely 64\,B.
With correct hints, \sysname takes 0.7\,\us to sign and 5.1\,\us to verify.
This is 27$\times$ and 7.0$\times$ faster than Dalek, respectively.
Interestingly, even though \sysname's larger signatures lead to a 1.0\,\us transmission overhead (more than 10$\times$ Dalek's), it has limited impact on its latency which is dominated by verification.
Overall, \sysname is 8.2$\times$ faster than Dalek.
With incorrect hints, \sysname's
signature verification
requires verifying both HBSS and EdDSA signatures, so verification latency increases to 39.9\,\us (4.3\,\us more than Dalek's).
Signature generation, however, is not impacted as signers still benefit from background EdDSA precomputation and the total latency, although rising to 41.5\,\us, is still 24\% lower than Dalek's.
Even with incorrect hints, \sysname has much lower combined sign-transmit-verify latency than the state of the art.

\subsection{Effect of Message Size on Latency}
\label{sec:msgsize}

We study the effect of message size on the
latency of \sysname by running the experiments of \S\ref{sec:evallat} with
  varying message sizes.

\begin{figure}
    \smallskip
    \centering
    \includegraphics[width=\columnwidth]{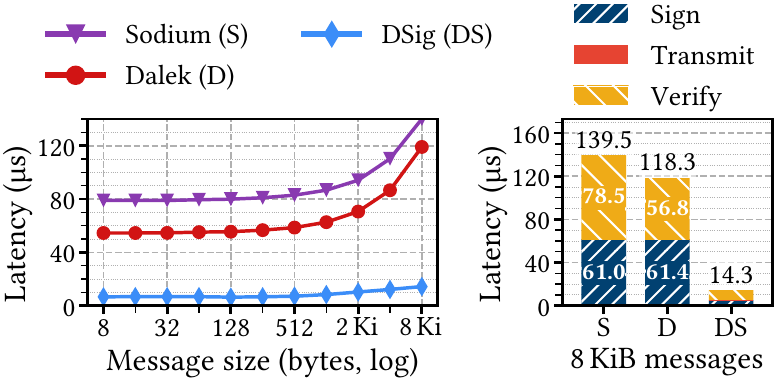}
    \caption{(Left) Latency to sign, transmit, and verify the signature of various-size
    messages using Sodium, Dalek and \sysname with correct hints.
    (Right) Median latency breakdown for 8\,KiB messages. Transmission overhead is invisible.}
    \label{fig:msg-size}
\end{figure}

Figure~\ref{fig:msg-size} (left) shows the results.
With larger messages, \sysname's total latency
  increases gradually but remains below 15\,\us.
Sodium's and Dalek's latencies are much higher.
They also increase faster
  because they use a slower hash function
  than \sysname (SHA256).\footnote{Most signature schemes
     hash the input to operate on a fixed-size string.}
Figure~\ref{fig:msg-size} (right) shows the
  latency breakdown for the largest size
  (the breakdown for the smallest size is
  in \S\ref{sec:evallat}).
In all schemes, the split is about half-half
  to sign and verify, with negligible
  transmission time.

\subsection{Throughput} \label{sec:eval-tput}

We study \sysname's throughput.
In an experiment, a process signs small 8\,B messages repeatedly
  with either a constant or an exponentially distributed random
  interval between signatures.
The signer transmits the signature over the network to the verifier,
  which verifies it.
We measure the total latency (sign plus transmit plus verify) and
  throughput of \sysname, and compare it against Sodium's and Dalek's.
In all experiments, the signer and the verifier use two cores each.
\sysname dedicates one core to each of its planes so that background events minimally impact foreground operations,
while Sodium and Dalek use all cores
to handle multiple messages in parallel.

\begin{figure}
    \smallskip
    \centering
    \includegraphics[width=\columnwidth]{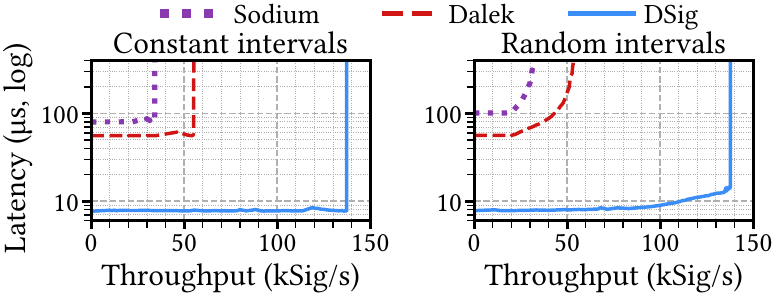}
    \caption{
    Latency-throughput graphs for Sodium, Dalek and \sysname.
    Signatures are issued at constant or exponentially
    distributed intervals.
    All three use two cores on both sides; \sysname dedicates one of them to its background plane.
    }
    \label{fig:tput-latency}
\end{figure}

Figure~\ref{fig:tput-latency} shows the results as
  latency-throughput graphs with median latency and
  average throughput.
With constant signature interval, all three systems exhibit stable latency until reaching maximum throughput.
Sodium maintains a latency of $\approx$80\,\us up to a throughput of 34\,kSig/s where it is bottlenecked by verification time (58\,\us).
Dalek maintains a latency of $\approx$56\,\us up to a throughput of 56\,kSig/s where it is also bottlenecked by verification time (36\,\us).
\sysname maintains a latency of $\approx$7.8\,\us up to a throughput of 137\,kSig/s where it is bottlenecked by the signer's background plane, which takes 7.4\,\us to generate a new public key.
We separately measured the verifier's background plane;
it achieves a throughput of 3.6\,MSig/s, so it is not a bottleneck.
With
a random signing interval,
  queuing occurs gradually, so the respective bottlenecks
  are less abrupt,
  causing a smoother latency degradation.

We run another experiment to measure the per-core throughput of \sysname by running both of its planes on one core, and compare it to the per-core throughput of Dalek.
While Dalek achieves 53\,kSig/s signature generations (resp. 28\,kSig/s verifications) per core,
\sysname achieves 131\,kSig/s signature generations (resp. 193\,kSig/s verifications) per core.

In summary, \sysname sustains significantly higher throughput at much lower latency than EdDSA-based systems.

\subsection{One-to-Many, Many-to-One Performance} \label{sec:eval-scalability}

We now study \sysname's scalability and bottlenecks in one-to-many and many-to-one scenarios.
In \emph{one-to-many}, one signer signs a message and sends
  the signature to many verifiers; this pattern is common
  in distributed protocols.
In \emph{many-to-one}, many signers sign different messages and send them to the same verifier; this pattern is common in client-server applications.
We run experiments
  where the
  signer(s) and verifier(s)
    use one foreground and one background core
    to work as fast as possible.
We measure the
  aggregate verification throughput, and report
  the average.
We consider a scenario where most of the network bandwidth
  ($\approx$90\%)
  is consumed by other activities, by
  limiting our NICs' bandwidth to 10\,Gbps.
This of course makes it harder for \sysname to operate
  since it consumes more network bandwidth than other
  schemes.
We compare the scalability of \sysname to a two-core system based on Dalek.

\begin{figure}
    \smallskip
    \centering
    \includegraphics[width=\columnwidth]{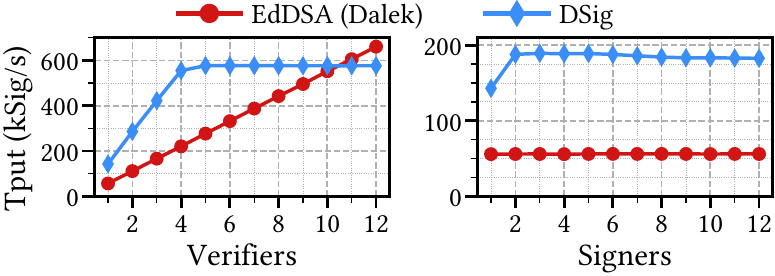}
    \caption{(Left) \sysname's throughput with a signer sending the same signature to multiple verifiers.
    (Right) \sysname's throughput with a verifier receiving different signatures from multiple signers.
    The NICs' bandwidth is limited to 10\,Gbps.}
    \label{fig:scalability}
\end{figure}

Figure~\ref{fig:scalability} shows the results.
In one-to-many (left of figure),
  \sysname's throughput increases until 577\,kSig/s with 5 verifiers;
  at this point, the signer
  saturates its link to the verifiers,
  with the 1,584\,B signatures and their 33\,B background data accounting for
  $\approx$7\,Gbps.
Dalek scales more slowly than \sysname with the number of verifiers, but it is not affected by bandwidth, as it continues to scale beyond 11 verifiers, at which point it surpasses \sysname's throughput with 603\,kSig/s using merely $\approx$300\,Mbps to transmit 64\,B signatures.

In many-to-one (right of figure),
  two signers are enough to
  achieve \sysname's maximum throughput of 190\,kSig/s
as they saturate the verifier's foreground plane, which we set to run on a single core.
As signing with Dalek is faster than verifying,
Dalek does not scale beyond 1 verifier and achieves a maximum throughput of 53\,kSig/s.

Overall, \sysname's main scalability bottleneck compared to Dalek is its larger signatures.

\subsection{Effect of Larger Signatures}\label{sec:eval-end-tput}

We study how \sysname's larger signatures affect application performance.
In each experiment, we run a synthetic application where a server receives signed requests of a given size,
 checks their signature,
  spends some given processing time, and sends a 16\,B unsigned reply.
Similarly to \S\ref{sec:eval-scalability}, we limit the NICs' bandwidth to 10\,Gbps, and compare the same application when using \sysname or EdDSA.
For fairness,
  EdDSA uses Dalek and pre-hashes
  messages with \blake.
In addition, we run an experiment with signatures disabled.
The application runs with 4 cores:
  \sysname uses 1 core for the background plane and
  3 cores to handle requests, while the others use
  4 cores to handle requests.
We run enough clients to saturate the server.
We consider 7 request sizes (32\,B, 128\,B, 512\,B, 2\,KiB, 8\,KiB, 32\,KiB and 128\,KiB) and
  2 processing times (1\,\us and 15\,\us).

\begin{figure}
    \smallskip
    \centering
    \includegraphics[width=\columnwidth]{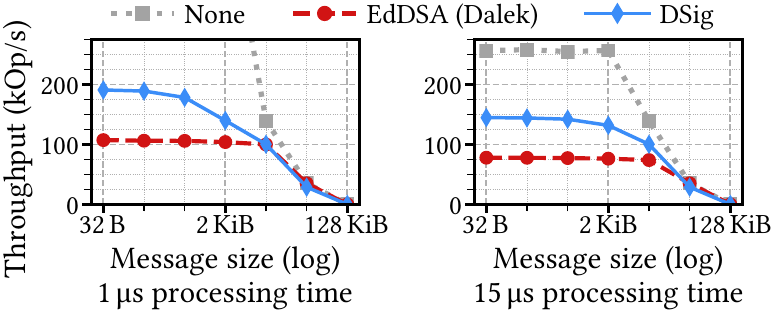}
    \caption{Request throughput of an application using
    signatures when
    NICs' bandwidth is constrained to 10\,Gbps,
    for different request sizes and
    request processing times.}
    \label{fig:bottleneck}
\end{figure}

Figure~\ref{fig:bottleneck} shows the results.
For both processing times, \sysname outperforms EdDSA up to
8\,KiB, after which it performs similarly to EdDSA.
For small messages (32\,B--512\,B), the limited bandwidth has no impact on
either scheme, so \sysname significantly outperforms EdDSA thanks to its lower computational cost.
With 2\,KiB messages and 1\,\us processing time, bandwidth impacts \sysname while EdDSA is almost unaffected.
Relative to 512\,B messages, \sysname's throughput decreases by 22\%, while EdDSA's decreases by only 1.9\%.
Higher processing time
offsets
\sysname's bandwidth bottleneck closer to 8\,KiB messages.
Beyond these points, the throughput of both \sysname and EdDSA converges to
that
of the application that does not use signatures, as network
bandwidth bottlenecks all three systems,
making the overhead of signatures negligible.

In summary,
  \sysname's higher per-core throughput lets applications reach higher throughput than with EdDSA
  even with limited network bandwidth, up to moderate-size messages.

\subsection{EdDSA Batch Size} \label{sec:eval-eddsa-batch}

To set the size of EdDSA-signed key batches (\S\ref{sec:eddsa-batching}),
we run the same experiment as in \S\ref{sec:evallat} for different batch sizes,
and we measure the latency and the per-core throughput.
To take into account the impact of larger batches on low-end networks, we limit our NICs' bandwidth to 10\,Gbps, as in
\S\ref{sec:eval-scalability}
and
\S\ref{sec:eval-end-tput}.

\begin{figure}
    \smallskip
    \centering
    \includegraphics[width=\columnwidth]{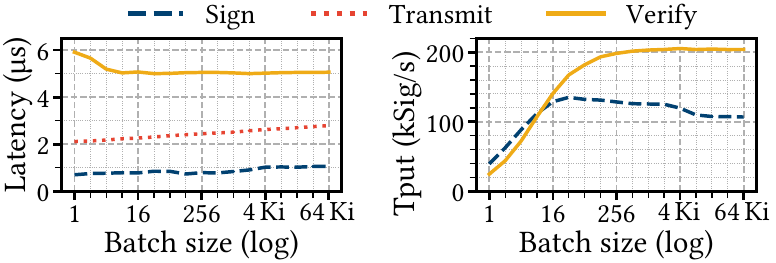}
    \caption{(Left) Median latency to sign, transmit, and verify a signature on an 8\,B message with \sysname for different EdDSA batch sizes. (Right) Single-core throughput of signing and verifying for different EdDSA batch sizes.}
    \label{fig:batch-size}
\end{figure}

Figure~\ref{fig:batch-size} shows the results, where
 a batch size of 1 means no batching.
We see that batch sizes do not affect latency much
  (left of figure).\footnote{
  The transmission latency differs from \S\ref{sec:evallat} due to the 10\,Gbps NIC limit.}
Throughput is different (right of figure):
  initially, batching improves throughput a lot
for both signing and verifying.
The gain dwindles when the amortized EdDSA cost per signature becomes a diminishing fraction of the overall computation as batches get larger.
The best signing throughput is 135\,kSig/s for batches of 32 keys,
while the best verifying throughput is 206\,kSig/s for batches of 4,096 keys.
We pick a batch size of 128 as a balance.

\section{Related Work} \label{sec:related}

\paragraph{HBSSs.}
HBSSs are well studied and prior work has proposed
  different implementations of them, many of
  which are variants of HORS~\cite{smart-grids, hors-ephemeral, horse, horsic-plus}.
Li \textit{et al.}~\cite{smart-grids} proposed a variant
  targeted at smart grids with limited
  storage
  that reduces key and signature size but increases
  computation costs.
Wang \textit{et al.}~\cite{hors-ephemeral}
proposed a scheme with small signatures and microsecond
  performance,
  but it is limited to providing
  low $\approx50$-bit security.
HORSE~\cite{horse} reduces the cost of few-time signatures
  by repeatedly hashing the private key secrets,
  creating a matrix whose last row is the public key; however,
  it restricts the order in which applications can reveal public keys.
\wots~\cite{wots-plus} was proposed by Hülsing as a variant of W-OTS~\cite{wots} with reduced signature and key sizes.

\paragraph{Online/offline signature schemes.}
The concept of online/offline digital signatures, in which heavy computation is done prior to knowing the message to sign, was first introduced by Even \textit{et al.}~\cite{even1990line}.
So far, practical applications of the theoretical concept (including hybrid signature schemes)  have targeted low-compute devices
and/or wide area networks,
with a focus on improving signature throughput or reducing bandwidth~\cite{onoff-lc, onoff-lp, onoff-wsn, onoff-wsn2, onoff-wsn3, onoff-wban}.
Recently, Esiner \textit{et al.}~\cite{lomos22} also recognized the importance of low-latency signatures,
yet their solution
is tailored for industrial control systems with tiny messages (25 bits), and does not provide self-standing signatures.
No prior work addresses hybrid signatures in data centers
  with microsecond-scale performance.

\paragraph{Merkle-based signatures.}
Prior work proposes schemes that rely exclusively on HBSSs to sign
(virtually) infinitely many messages, with the goal of
attaining quantum resistance.
Most of this work is based on XMSS~\cite{xmss}, such as SPHINCS~\cite{sphincs}
and variants~\cite{sphincs-plus, sphincs-plus-c, xmss-plus}.
Instead of distributing keys regularly,
  these schemes efficiently pack an
  infinitude of one-time public keys using Merkle inclusion proofs~\cite{mss}.
  These proofs need to be checked during signature verification,
thus making the performance of such schemes be in the milliseconds.

\paragraph{Signature-like schemes.}
The cost of signatures has fueled alternatives for
  different scenarios.
Message authentication codes (MACs) provide authentication and integrity of messages,
but lack transferability, as parties use a shared secret to communicate.
While MACs are widely used in networked systems to provide
  authenticated channels between two parties, they
  are not substitutes for signatures, as they provide
  weaker properties, they are harder to use, and they
  are more susceptible to protocol mistakes.
In particular, using MAC-based mechanisms in BFT protocols has several drawbacks:
(1) These mechanisms are ad-hoc and highly dependent on the protocol: some require MAC vectors~\cite{pbft-macs} others require MAC matrices~\cite{matrix-signatures}; others explicitly prefer or mandate signatures over MACs for critical messages~\cite{aguileraBGPXZ21,ubft,aardvark}.
(2) These mechanisms add complexity to the BFT protocols, e.g., by requiring a fast-slow path approach where the fast path avoids signatures but the slow path (or view change) still uses them~\cite{pbft-macs,ubft,aguileraBGPXZ21,zyzzyva,aardvark,hq};
this added complexity increases their attack surface~\cite{aardvark}.
(3) These mechanisms often add messages and roundtrips to the protocols~\cite{matrix-signatures, tesla, hq}, and/or lower their resilience to failures~\cite{matrix-signatures, tesla}.

Some systems make extra assumptions to
    provide MAC with some form of transferability.
TESLA~\cite{tesla} assumes clock synchrony and
  has time windows during which MACs are generated and
  transmitted; afterward, the MAC secrets are
  revealed to check previously seen MACs.
This idea provides only a limited form of
  transferability and increases the latency of verification.
Using trusted hardware~\cite{levin2009trinc,KapitzaBCDKMSS12,behl2017hybrids}, such as trusted execution environments (TEEs), one can
provide MACs with transferability by hiding the
secret and computing the MACs in the TEE so that every
TEE owner can verify the MACs but only a designated TEE can create them.

\section{Conclusion} \label{sec:conclusion}

\sysname is the first digital signature system for microsecond-scale applications.
\sysname
  achieves single-digit microsecond latency for
  signing and verifying
  messages---27${\times}$ and 7${\times}$ faster
  than the prior state of the art---while achieving higher throughput.
To achieve that, \sysname introduces a new hybrid signature scheme
that uses knowledge of where signatures are issued and verified
  in the common case.
\sysname can bring auditability to latency-critical applications with a small latency overhead,
or replace other signature schemes in applications that use them.
Ultimately, we believe that \sysname makes digital signatures fast enough to
broaden their use in data centers
as a powerful security building block.

\section*{Acknowledgments} \label{sec:acknowledgments}
We thank the anonymous reviewers and our shepherd Steve Hand
for their valuable comments,
as well as the anonymous artifact evaluators for reviewing our implementation.
We also thank Dariia Kharytonova, Ed Bugnion, Jean-Philippe Aumasson, Khashayar Barooti, Phillip Gajland, Pierre-Louis Roman, and Serge Vaudenay for their feedback.

\newpage

\bibliographystyle{plain}
\bibliography{references}

\end{document}